\documentclass[a4,12pt,epsf]{article}
\usepackage{graphicx}
\begin{document}
\begin{center}{\Large \bf
Dark Matter Mass and Anisotropy \\
in Directional Detector
}
\end{center}
\begin{center}
Keiko I. Nagao
\footnote{nagao@dap.ous.ac.jp}
\footnote{This talk is based on the paper \cite{Nagao:2017yil}.}
\vspace{6pt}\\
{\it
Okayama University of Science, 
1-1 Ridaicho, Kita-ku, Okayama \\700-0005 Japan
}
\end{center}
\begin{abstract}
Velocity distribution of dark matter is supposed to be isotropic Maxwell-Boltzmann distribution in most cases, however, other distribution models including anisotropic one are suggested by simulations. Directional direct detection of dark matter is expected to be a hopeful way to discriminate isotropic distribution from anisotropic one. We investigate the possibility to obtain the anisotropy and WIMP mass by Monte-Carlo simulation supposing the directional detector.
\end{abstract}

\section{Introduction}
Cosmological and astrophysical observations have showed that dark matter consists 27\% of the energy density of the Universe,
which corresponds to about 5 times as much as baryonic matter. 
Weakly interacting massive particles (WIMPs) are a hopeful candidate of dark matter.
In spite of many projects to hunt dark matter such as direct, indirect detections and collider search, we still know very little about the dark matter. Velocity distribution is one of essential quantity of dark matter. Especially, it can affect derivation of constraints for dark matter mass and cross section in the direct detection. 
Direct detection which has directional sensitivity is hopeful experiment to reach the local velocity distribution of WIMPs \cite{Mayet:2016zxu}. 
Especially, it is suitable to investigate anisotropic components of the velocity distribution suggested by N-body simulation
\cite{LNAT}. 
In the reference, the velocity distribution associated with tangential velocity $v_\phi$ in the Galactic rest frame is indicated as 
\begin{eqnarray}
f(v_\phi) = \frac{1-r}{N(v_{0,\mathrm{iso.}})} \exp\left[-v_\phi^2/v_{0,\mathrm{iso.}}^2\right] + 
		\frac{r}{N(v_{0,\mathrm{ani.}})} \exp\left[-(v_\phi-\mu)^2/v_{0,\mathrm{ani.}}^2\right] ,
\label{eq:doublegaussian}
\end{eqnarray}
where $N(v_{0,\mathrm{iso.}})$ and $N(v_{0,\mathrm{ani.}})$ are normalization factors,  $r=0.25$ is anisotropic parameter, $v_{0,\mathrm{iso.}} = 250$ km/s, 
$v_{0,\mathrm{ani.}} = 120$ km/s and $\mu=150$ km/s. Note that the velocity distribution is isotropic if $r=0.00$. 
We investigate the possibility to discriminate anisotropic velocity distribution Eq. (\ref{eq:doublegaussian}) from isotropic distribution by the directional direct detections including a solid type detector.

\section{Detection in Directional Detector}
\begin{figure}[ht] 
   \centering
   \includegraphics[width=2in]{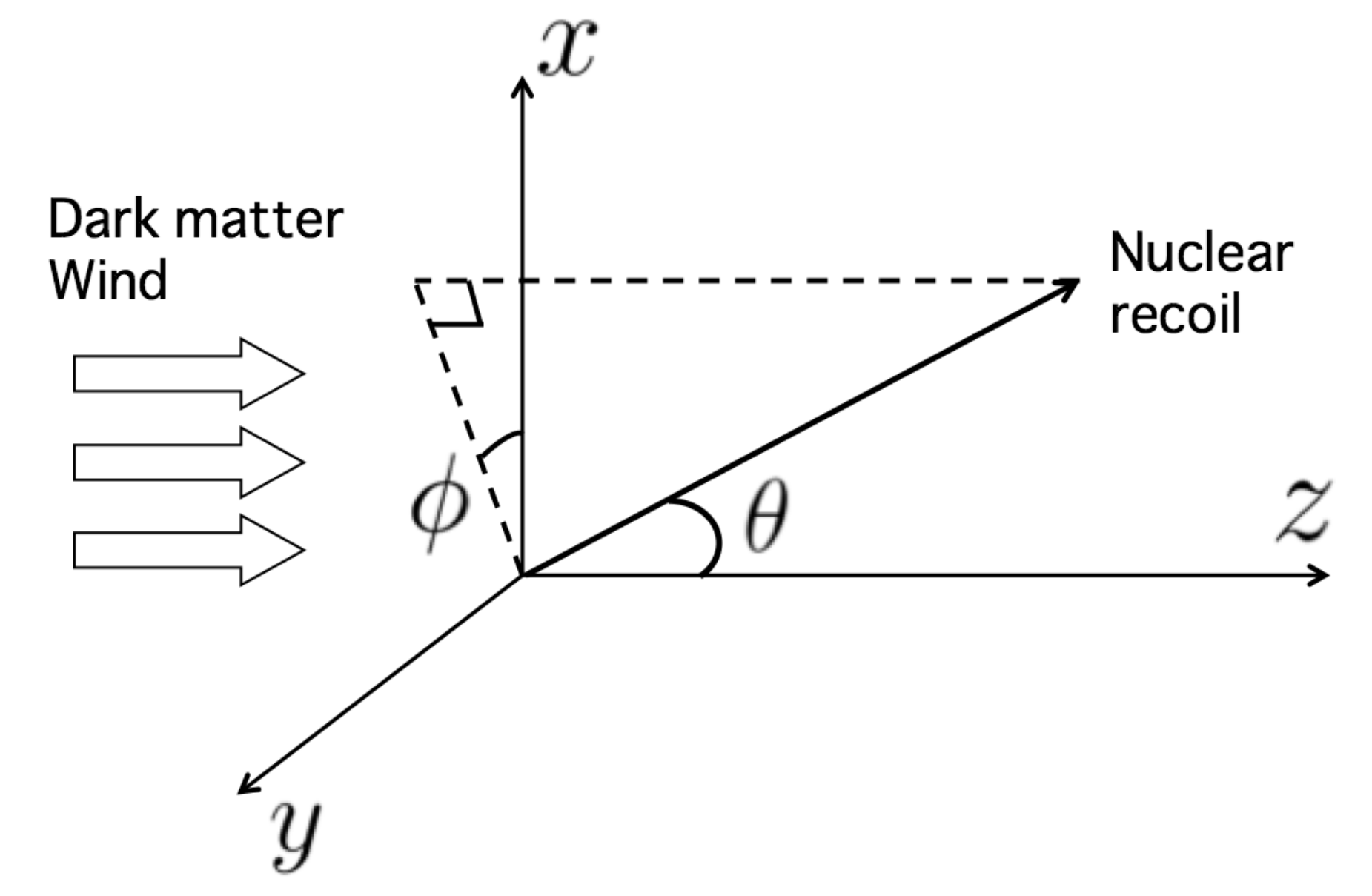} 
   \caption{WIMP-nucleon scattering in the Galactic rest frame.}
   \label{fig:labsys}
\end{figure}
In Figure \ref{fig:labsys}, WIMP - nucleon scattering in the directional detector is shown. The z-axis is taken as the direction of WIMP wind towards the Solar system. Supposing the velocity distribution of corresponding anisotropy $r$, a WIMP is generated following 
a probability of the distribution and the recoil energy and the scattering angle are obtained in the Monte-Carlo simulation. 
Most of the directional direct detector are gas detector in which fluorine (F) is adopted as a target atom. In the solid directional detector, there are several target atoms including silver (Ag), bromine and carbon. Among them we suppose F and Ag as target atoms in the numerical simulation. 

In order to distinguish the energy-angular distribution obtained by supposing anisotropic velocity distribution from that by supposing anisotropic one, two kinds of data is generated by the Monte-Carlo simulation. 
 
\section{Numerical Result}
\subsection{Case 1 : Supposing the WIMP mass is known.}
\begin{figure}[t]
 \centering
  \includegraphics[width=5cm,clip]{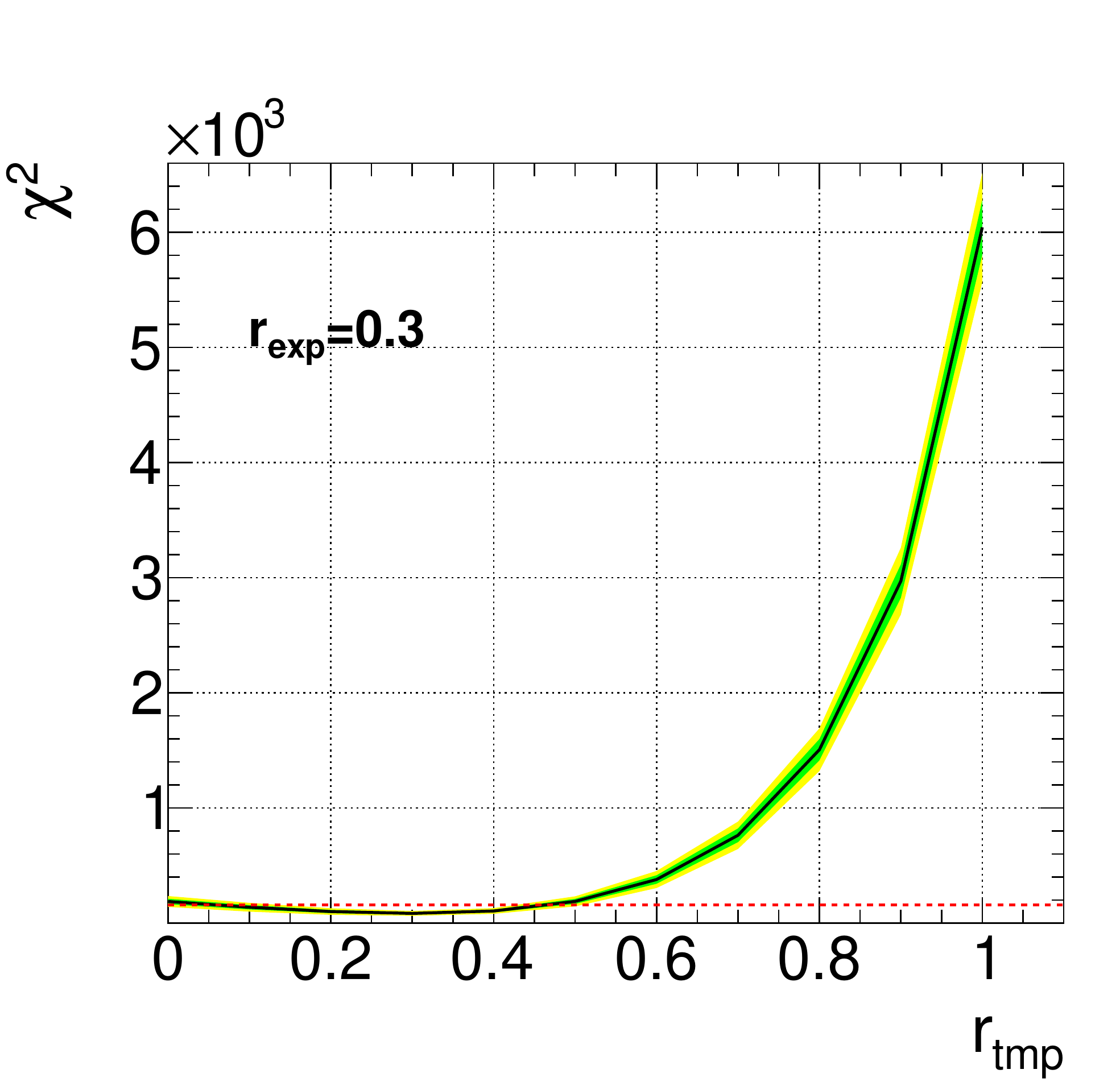}
    \includegraphics[width=5cm,clip]{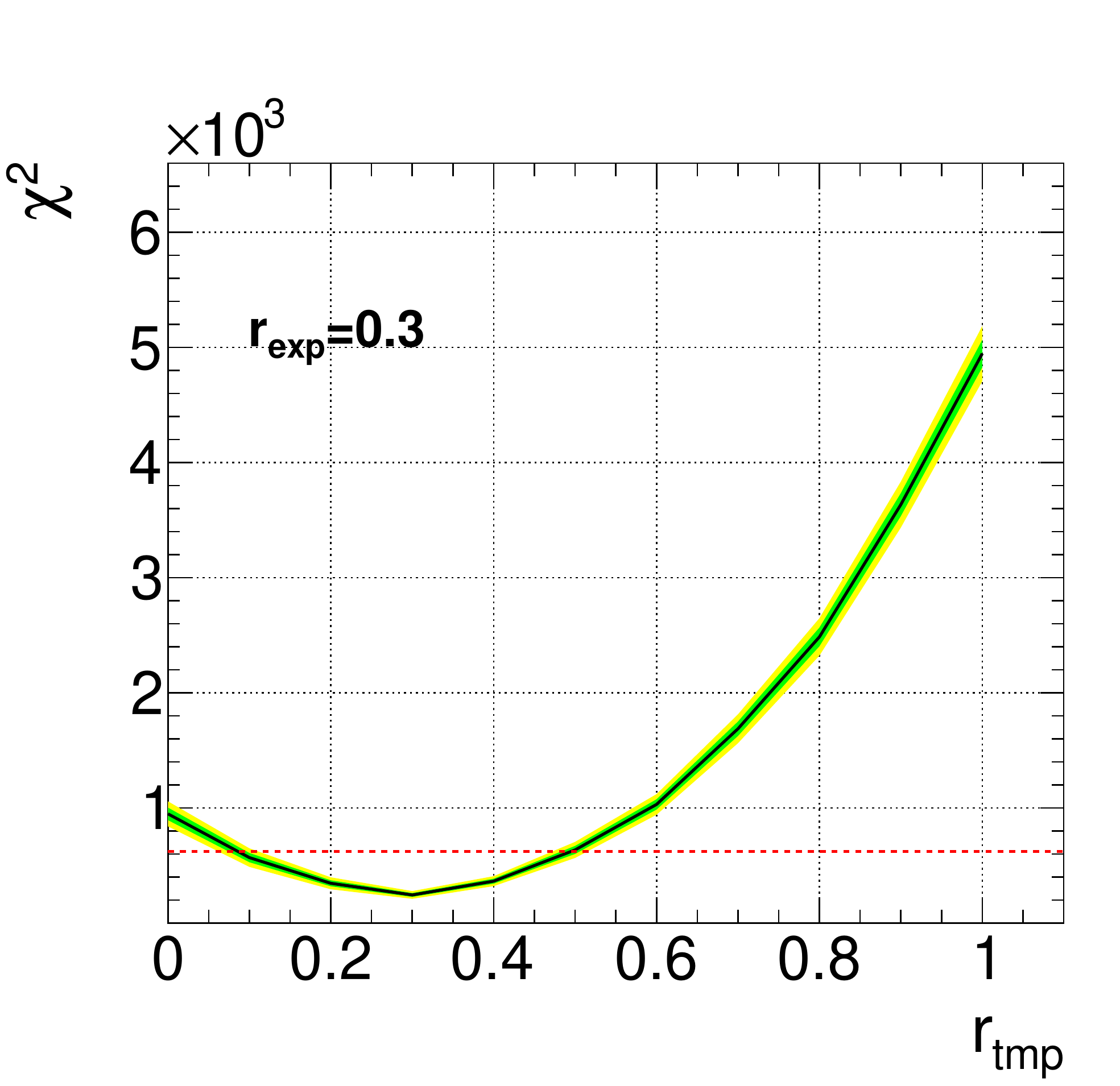}
 \caption{(Left) The chi-squared test of the energy-angular distributions for target F. Red dotted lines show 90\% C.L. The event number of pseudo-experiment is set as 6$\times 10^3$, and the energy threshold of the detector is 20 KeV. (Right) The chi-squared test for target Ag case. The event number of pseudo-experiment is set as 6$\times 10^4$, and the energy threshold of the detector is 50 KeV.}
 \label{fig:chi2_ERnon0}
\end{figure}
In the case that the WIMP mass is obtained by other experiment, the energy threshold can be optimized to distinguish the anisotropy of the velocity distribution.
In Figure \ref{fig:chi2_ERnon0}, the chi-squared test for target F and Ag are shown. If the anisotropic case is realized, the required event number is 6$\times 10^3$ (target F) and 6$\times 10^4$ (target Ag) in order to reject the isotropic distribution by 90 \% confidence level (C.L.).

\subsection{Case 2 : Constraining both WIMP mass and velocity distribution.}
\begin{figure}[ht] 
 \centering 
 \includegraphics[keepaspectratio, scale=0.19,clip]{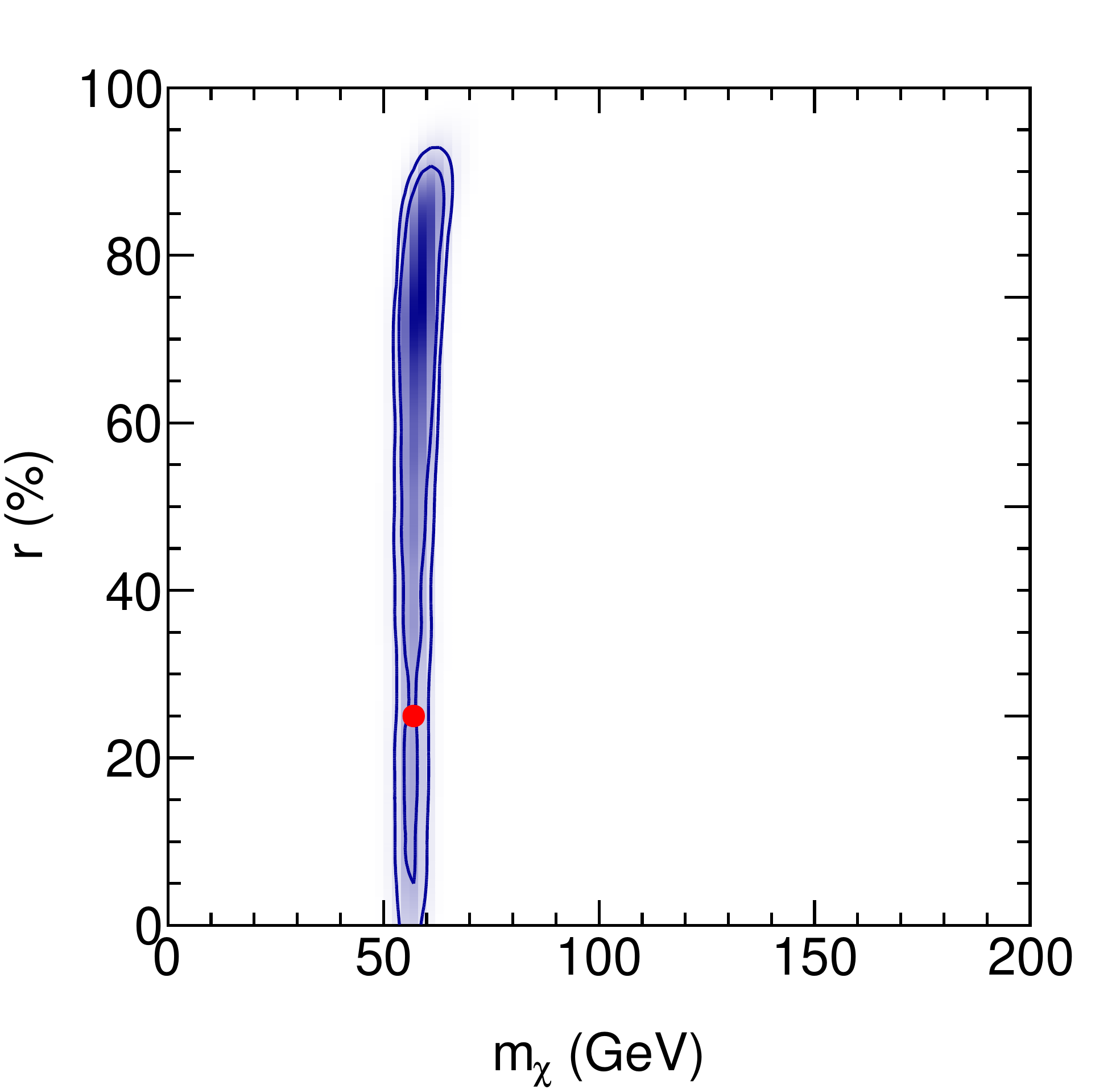} 
\includegraphics[keepaspectratio, scale=0.19,clip]{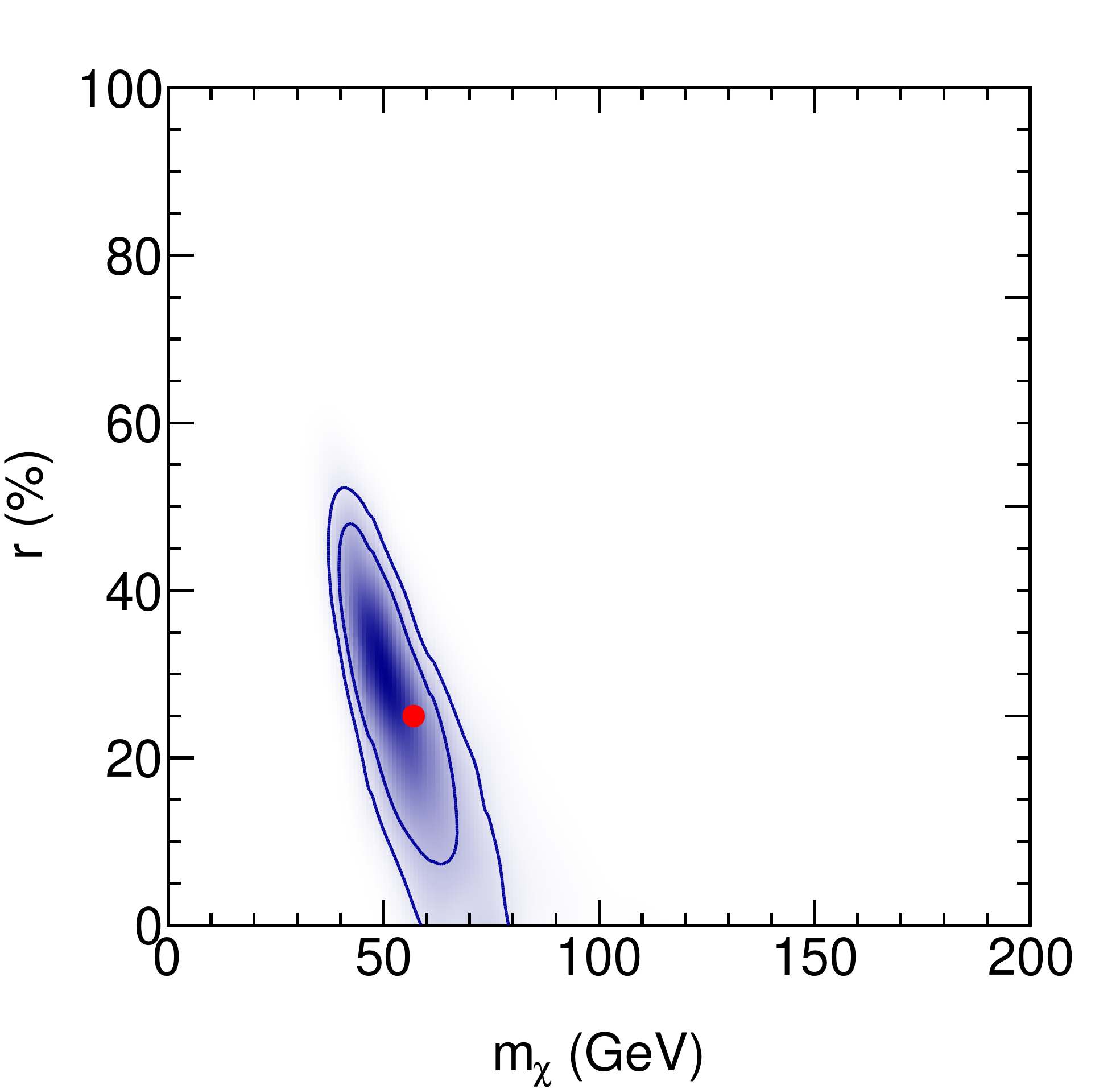} 
\includegraphics[keepaspectratio, scale=0.19,clip]{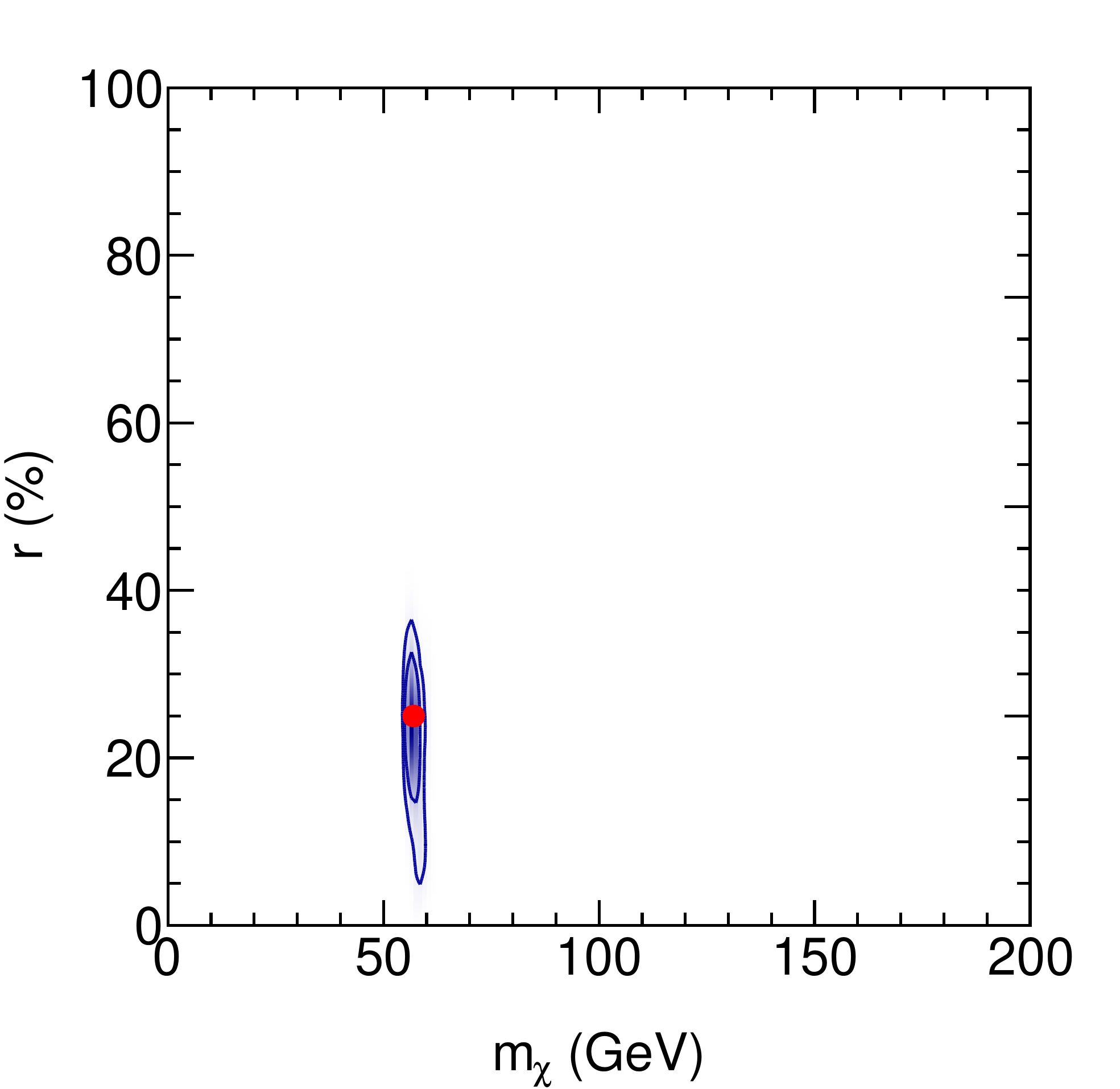} 
 \caption{The 2D posterior probability distributions in the WIMP mass and anisotropy space for target F.  
 Left: Only data of recoil energy $E_R$ is used. Center: Only data of scattering angle $\cos{\theta}$ is used. Right: Both recoil energy and scattering angle are used.}
 \label{fig:likelihoodF}
\end{figure}

\begin{figure}[ht]
 \centering
 \includegraphics[keepaspectratio, scale=0.19,clip]{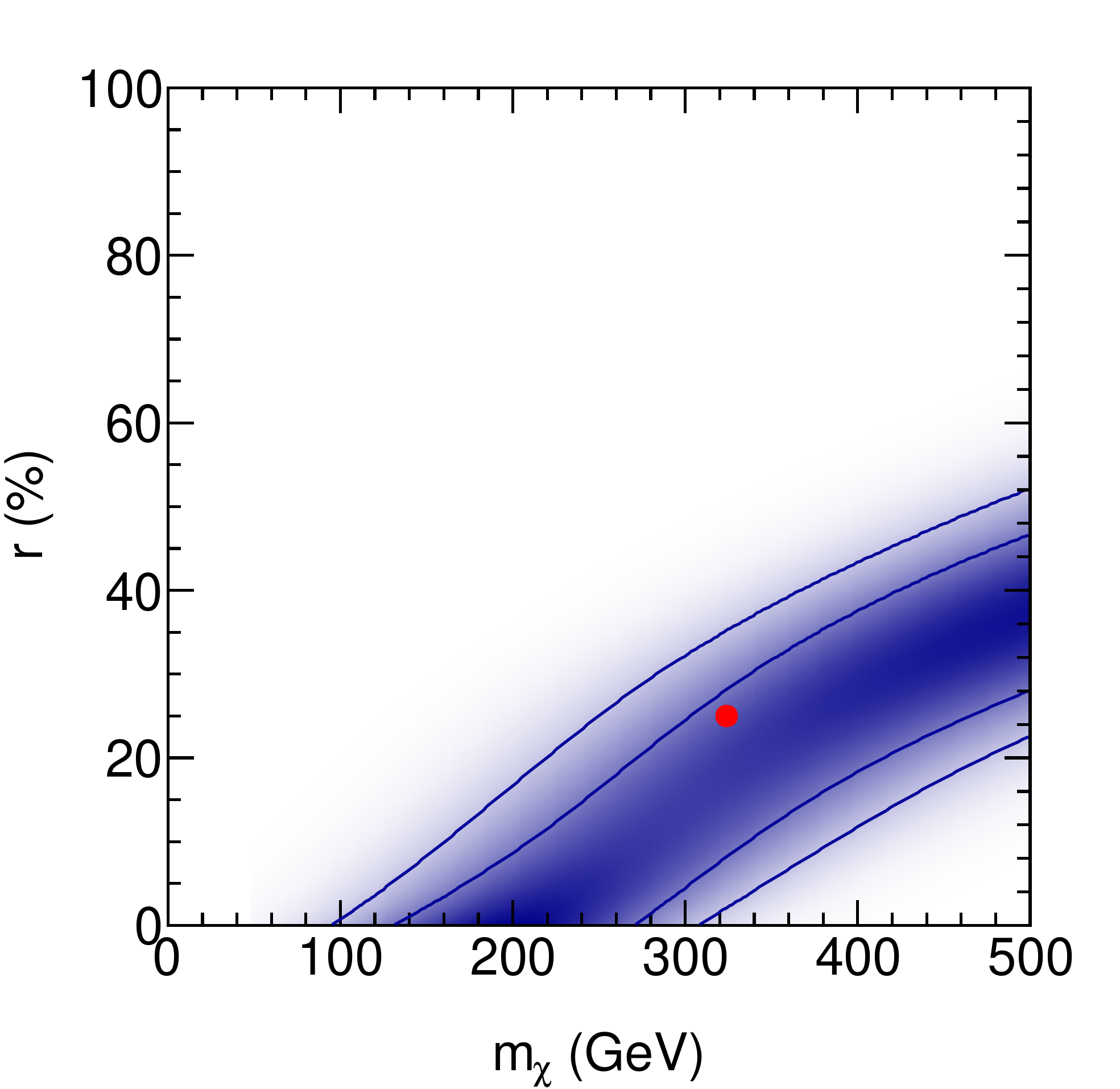}
\includegraphics[keepaspectratio, scale=0.19,clip]{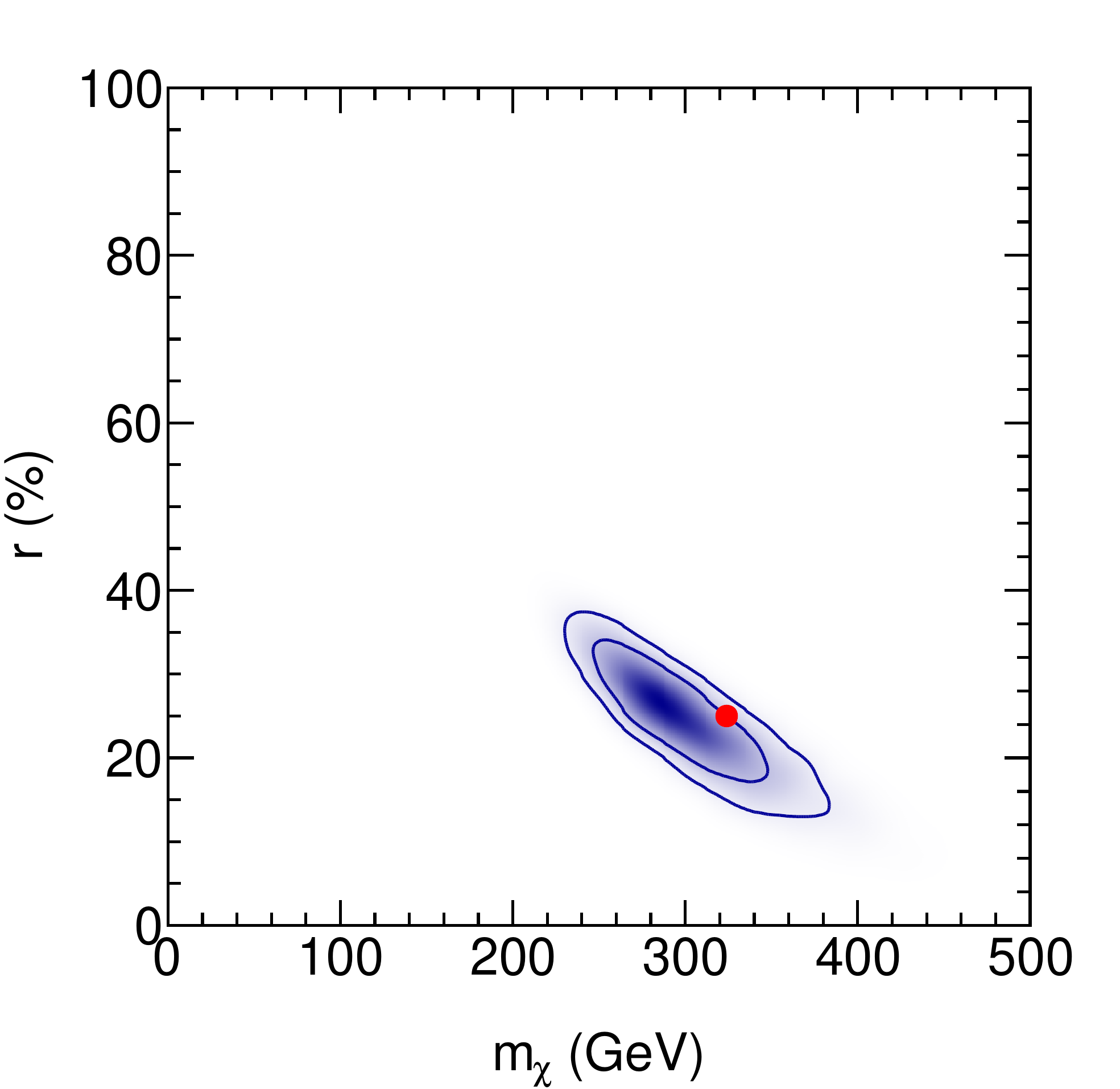}
\includegraphics[keepaspectratio, scale=0.19,clip]{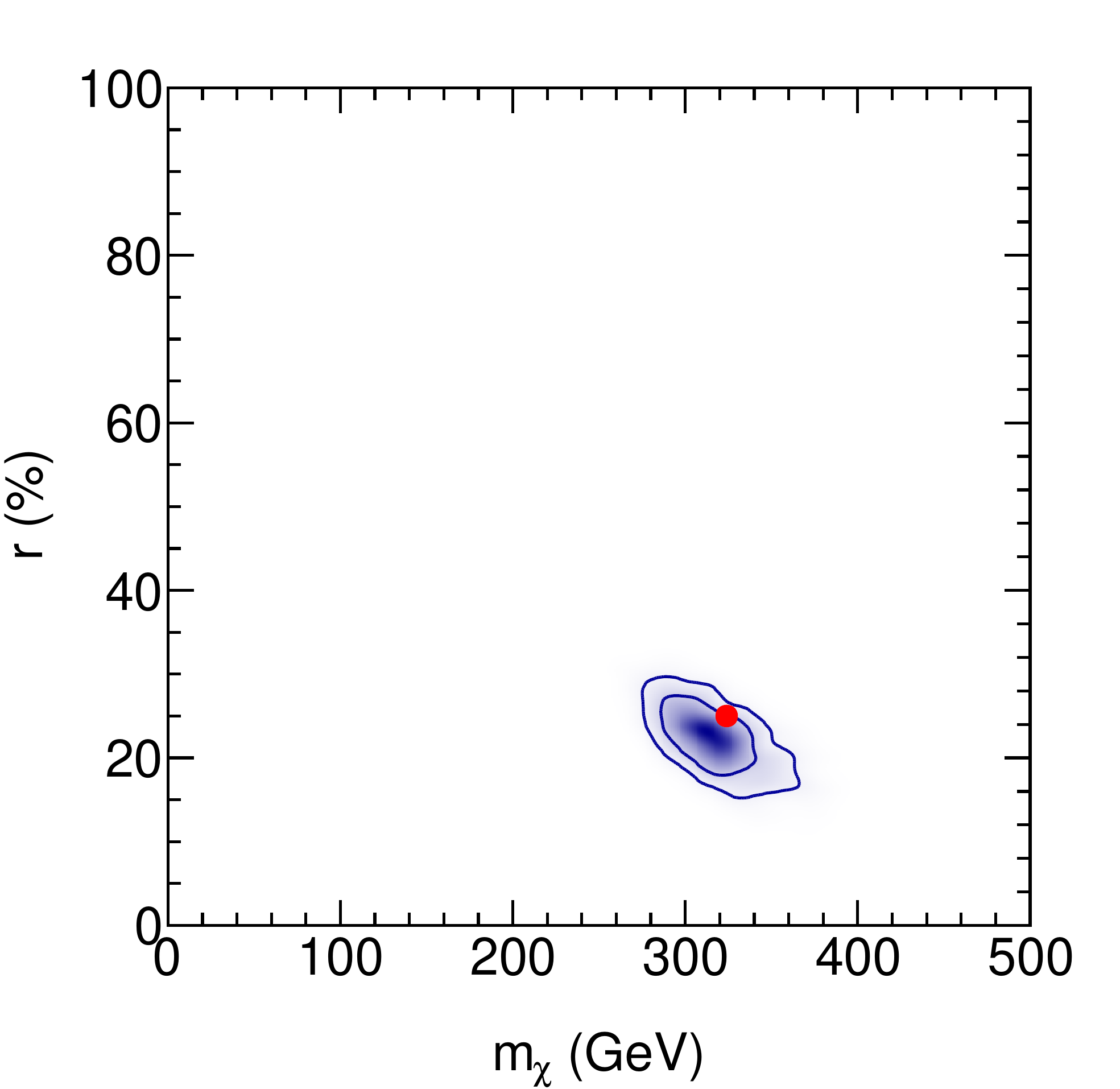}
 \caption{Legend is same as Figure \ref{fig:likelihoodF} but for the target Ag.}
 \label{fig:likelihoodAg}
\end{figure}

Even if information of WIMP mass is not provided, a constraint for both WIMP mass and anisotropy can be obtained. In Figure \ref{fig:likelihoodF} and \ref{fig:likelihoodAg}, the probability distribution for target F and target Ag obtained by likelihood method are shown, respectively. With only directional data or only the recoil energy data, indication for the parameters is unreliable compared to the case that both data are used in the analysis.


\section{Summary}
Discrimination of the anisotropy of the WIMP velocity distribution in the directional detector is investigated. Once the WIMP mass is obtained, supposing 30\% of anisotropy component of the velocity distribution, the required event numbers for discriminate the anisotropy are O(1000) for light target and O(10000) for heavy one, respectively. Even without the WIMP mass information, the WIMP mass and anisotropy can be obtained with less uncertainty by using both the recoil energy and the scattering angle data, than in the analysis by using only one of the data.
\vspace*{12pt}
\noindent
\\ 


\end{document}